\documentstyle[11pt]{article}

\def\ba{\begin{eqnarray}}
\def\ea{\end{eqnarray}}
\def\be{\begin{equation}}
\def\ee{\end{equation}}
\newfont{\msbm}{msbm10}
\newfont{\msbms}{msbm6}  
\newfont{\cmss}{cmss10}  
\def\g{\gamma}

\def\C{{\cal C}}

\def\P{{\cal P}}

\def\A{{\cal A}}

\def\H{{\cal H}}

\def\ag{{{\cal A}/{\cal G}}}
\def\agb{{\overline \ag}}

\def\Lagm0{L^2\bigl(\agb,\m_0\bigr)}

\def\h0{$\H_0$}

\def\Ab{\mbox{$\overline \A$}}
\def\ca{\mbox{$C(\Ab)$}}

\def\m{\mu}

\def\l{\lambda}

\def\o{\omega}

\def\rep{representation}
\def\reps{representations}
\def\N{\hbox{\msbm N}}

\def\R{\hbox{\msbm R}}

\def\CC{\hbox{\msbm C}}
\begin{document}

%%%%%%%%%%%%%%%%%%%%%%%%%%%%%%
%\baselineskip=18pt
%%%%%%%%%%%%%%%%%%%%%%%%%%%%%

%%%%%%%%%%%%%%%%%%%%%%%%%%%%%%%%%%%%%%%%%%%%%%%%%%%%%%%%%%%%%%%%%%%%%%

\title{Denseness of Ashtekar-Lewandowski  states\\
and  a generalized cut-off in loop quantum gravity}

\author{J. M. Velhinho}

\date{{Departamento de F\'\i sica, Universidade da Beira 
Interior\\R. Marqu\^es D'\'Avila e Bolama, 
6201-001 Covilh\~a, Portugal}\\jvelhi@ubi.pt}

\maketitle
%%%%%%%%%%%%%%%%%%%%%%%%%%%%%%%%%%%%%%%%%%%%%%%%%%%%%%%%%%%%%%%%%

%%%%%%%%%%%%%%%%%%%%%%%%%%%%%%%%%%%%%%%%%%%%%
\begin{abstract}

\noindent  
We show that the set of states of the Ashtekar-Isham-Lewandowski holonomy algebra defined by elements
of the Ashtekar-Lewandowski Hilbert space is  dense in the space of all states. 
We consider weak convergence properties of a 
%possible generalization
modified version of the 
%standard 
cut-off procedure currently in use in loop quantum gravity.
%to produce normalizable 
%(graph-dependent) 
%states out of distributional ones. 
This version  is  adapted to vector states rather than to general distributions.
%to true states of the algebra that are not realizable as elements of the 
%Ashtekar-Lewandowski Hilbert space.
\end{abstract}
%%%%%%%%%%%%%%%%%%%%%%%%%%%%%%%%%%%%%%%%%%%%%%
%%%%%%%%%%%%%%%%%%%%%%%%%%%%%%%%%%%%%%%%%%%%

%%%%%%%%%
%\newpage 
%%%%%%%%%%%%%

%%%%%%%%%%%%%%%%%%%%%%%%%%%%%%%%%%%%%%%%%%%%
\pagestyle{myheadings}
%\markboth{J. M. Velhinho}{Ashtekar-Lewandowski  states in loop quantum gravity}
%\markright{
%%%%%%%%%%%%%%%%%%%%%%%%%%%%%%%%%%%%%%%%%%%%%%%%%%%%%%%%%%%%%%%%%%%%%%%%%%%%

%%%%%%%%%%%%%%%%%%%%% BEGIN PAPER %%%%%%%%%%%%%%%%%%%%%%%%%

%%%%%%%%%%%%%%%%%%%%%%%%%%%%%  Section  %%%%%%%%%%%%%%%%%%%%%%%%%%%%%%%%%
\section{Introduction and overview}
\label{int}
%%%%%%%%%%%%%%%%%%%%%%%%%%%%%%%%%%%%%%%%%%%%%%%%%%%%%%%%%%%%%%%%%%%%%%%%%
The  kinematical algebra on which loop quantum gravity (LQG) is based,
namely the algebra of holonomies and fluxes, has a preferred,
seemingly unique, 
\rep\ defined by a state, or measure $\m_0$. This state gives rise to the so-called
kinematical Hilbert space, a non-separable space $\H_{0}:=L^2(\Ab,\m_0)$ of square integrable
functions on the compact space \Ab\ of generalized connections (see \cite{AL5,R,T} for reviews of
LQG and \cite{F,LOST,OL,OL2,ST1} for recent uniqueness results concerning the $\H_0$ \rep). 
Nevertheless,  different states and corresponding \reps\ of the holonomy
$C^*$-algebra $C(\Ab)$ -- the commutative sub-algebra of the full kinematical algebra corresponding 
to configuration variables -- are still worth considering, in particular in relation to 
the active search for low energy
or semiclassical states~\cite{AG,AL4,Bo,T,T1,Va1,Va2,Va3,Ve}.

In the present work we address the question of completeness of the Hilbert space $\H_0$ 
with respect to the space of all states of the holonomy algebra $C(\Ab)$, in the following sense.
Let us consider the set $E(\H_0)$ of states of $C(\Ab)$ defined by normalized vectors $\psi\in\H_0$, i.e.~states 
of the form 
$$\ca\ni f \mapsto \langle\psi,f\psi\rangle_0,$$
where $\langle\ ,\rangle_0$ denotes the $\H_0$ inner product. 
The first question we wish to address is whether an arbitrary state
of \ca\ can be approximated, or obtained in some limit, from elements of $E(\H_0)$.
A variant of this question has been considered in~\cite{T1}, where it was argued that
 new \reps\ can indeed be obtained from \h0\ through standard limiting procedures.
Partly motivated by that work, we aim at a rigorous formulation of completeness results
of the above type. Specifically, we consider the natural weak-$\star$ topology on the space of states
of \ca\ and present a proof of the denseness of the set $E(\H_0)$. 
%%%%%
%As we will see, this follows
%directly from the faithfulness of the \h0\ \rep. 
%%%%%%%%%%%
%Such a result was actually to be expected, or,
%to put it differently, can be seen as an application of well known facts from the theory of
%commutative $C^*$-algebras, or measure theory in compact Hausdorff spaces.

A second, more subtle issue  is inspired by the so-called "cut-off" procedure,
which is a consistent way to deal with formal, non-normalizable linear combinations of  $\H_0$
vectors that arise naturally in the context of LQG semiclassical analysis~\cite{AL4,T1}. 
In broad terms, to be refined below,
%the  question 
we want to consider 
%is 
how to map a given state of \ca, and elements of the corresponding
Hilbert space, to nets of $\H_0$ vectors, preserving as much information as possible.
%%%%%%%%%%%%%
%To begin with, this is a problem of finding a net  $\{\psi_{\l}\}_{\l\in I}$,
%where $I$ is some index set and  $\psi_{\l}\in\H_0$, $\forall \l \in I$, such that the net 
%$\{\langle \psi_{\l},\Pi(\cdot)\psi_{\l}\rangle_0\}_{\l\in I}$ converges to a given state
%$\omega(\cdot)$. 
It turns out that that the projective structure built in the framework 
of LQG
provides, as in the cut-off procedure, the key
ingredients to address this question.

As discussed in more detail in section \ref{lp}, the space \Ab\ is a projective limit of a family of 
finite dimensional spaces $\A_\g$, each one associated to a graph $\g$. Moreover, a state of \ca,
or a measure $\m$ in \Ab, is equivalent to a family of measures $\m_\g$, each defined in the corresponding
space  $\A_\g$. Typically for measures $\m$ of potential interest one expects that, while the Radon-Nikodym 
derivative $d\m/d\m_0$ fails to exist, the corresponding derivatives $d\m_\g/d{\m_0}_\g$ remain
well defined. It follows that, for every measure as above, one can naturally define a map
$\psi\mapsto \{\psi_{\g}\}$ from $\H=L^2(\Ab,\m)$ to nets of \h0\ vectors. As in the cut-off procedure,
these nets are labeled by graphs and each $\psi_\g$ is (the pull-back of) an element of
$L^2(\A_\g,{\m_0}_\g)$. 
We consider here convergence 
properties of this map. 
Besides 
convergence of expectation values of elements of \ca, we will see
that convergence of arbitrary matrix elements is also achieved, 
i.e.~$\langle \psi'_{\g},f\psi_{\g}\rangle_0$ 
converges to
$\langle \psi',f\psi\rangle$, $\forall\psi',\psi\in\H$, $\forall f\in C(\Ab)$.
The map $\psi\mapsto \{\psi_\g\}$ can be seen as a (weak) generalization of the would be
transformation
%%%%%%%
%\be
%\label{int0}
$$ \H\ni\psi\mapsto (d\m/d\m_0)^{1/2}\psi\in\H_0, $$
%\ee
%%%%%%%%%
which, of course, depends on the existence of
$d\m/d\m_0$.
%%%%%%%%%%
%These properties reinforces the status
%of the map  $\psi\mapsto \{\psi_\g\}$ as an appropriate weak replacement of the
%non existing transformation (\ref{int0}), suggesting that it could be viewed,
%in the cases under consideration, as a canonical way of incorporating elements
%of different \rep\ Hilbert spaces into \h0.

The questions considered in the present work have well defined, and certainly well
studied analogues in standard quantum field theory (QFT). The reasons to reconsider
them in the LQG context are twofold. First, QFT typically deals with infinite dimensional 
linear spaces, and measures that give rise to familiar separable Hilbert spaces.
It is thus not obvious what kind of results have straightforward analogues in LQG,
where one finds an infinite dimensional compact space \Ab, and a fiducial Hilbert space $\H_0$
which is non-separable. 
%Moreover, experience with non-separable Hilbert spaces in physics
%is scarce, and the communication of corresponding results to a physicists audience is,
%we believe, useful, even if the results turn out to be as expected.   

Second, the status of the questions we consider is likely to be  different
in QFT and in LQG. To begin with, and still deeply related to the first point,
in QFT one does not have a true analogue of the $\m_0$
measure, as this would have to be the non-existing "Lebesgue measure on an infinite
dimensional linear space". Gaussian measures take its place as fiducial measures,
but these are not as distinguished, or singled out in the way $\m_0$ is.  
Moreover, there are specific reasons why maps between inequivalent \rep\ Hilbert spaces,
such as the map $\psi\mapsto \{\psi_{\g}\}$ above, may be useful in LQG, while
they can be regarded as constructions of disputable interest in QFT. Let us point out
that the convergence $\langle\psi'_\g,\cdot\,\psi_\g\rangle_0\to\langle\psi',\cdot\,\psi\rangle$
one obtains is usually far from being uniform. Thus, typically no fixed $\psi_\g$ 
approximates well the properties of $\psi$ with respect to a  large set of operators.
This diminishes the interest of such maps when, as in QFT, the operators in question
are physical observables. In LQG, however, variables like $f\in\ca$ are not observables
(as they do not commute with the constraints of general relativity).
There is thus hope that  images $\psi_\g$ can approximate the properties of corresponding $\psi$
with respect to a more meaningful restricted set of operators, constructed e.g.~from elements
of \ca. That this is a real possibility, regarding e.g.~a sufficient set of spatially
diffeomorphism invariant operators, is one of the conclusions of~\cite{T1}.

This work is organized as follows. In section \ref{pre} we review selected mathematical aspects of 
the LQG approach. In section \ref{newgen} the denseness of the set of states
$E(\H_0)$ is proved.
In section \ref{pro}  the above mentioned convergence properties
of the nets $\{\psi_\g\}$ are proved.
%%%%%%%%%%%%%%%%%%%%%%%%%%%%%  Section  %%%%%%%%%%%%%%%%%%%%%%%%%%%%%%%%%
\section{Preliminaries}
\label{pre}
%%%%%%%%%%%%%%%%%%%%%%%%%%%%%%%%%%%%%%%%%%%%%%%%%%%%%%%%%%%%%%%%%%%%%%%%%
In this section we review very briefly  the necessary mathematical 
aspects of LQG (see e.g.~\cite{AL5,T,T2,Ven} for a more thorough review). 
We also include a 
summary of notions and 
results used throughout the present work, mainly topics from the theory of
commutative $C^*$-algebras (see e.g.~\cite{BR} and~\cite{RS} for an extensive coverage).
%%%%%%%%%%%%%%%%%%%%%%%%%%% 
%%%%%%%%%%%%%%%%%%%%%%%%%%%%%  Section  %%%%%%%%%%%%%%%%%%%%%%%%%%%%%%%%%
\subsection{Summary of commutative $C^*$-algebra results}
\label{gen}
%%%%%%%%%%%%%%%%%%%%%%%%%%%%%%%%%%%%%%%%%%%%%%%%%%%%%%%%%%%%%%%%%%%%%%%%%
Let $\C$ be a commutative $C^*$-algebra with identity. There exists a uniquely defined compact Hausdorff 
space $X$ -- called the spectrum of $\C$ --
such that $\C$ is 
isomorphic to the algebra $C(X)$ of complex
continuous functions on $X$. One can therefore identify $\C$ with $C(X)$.

%%%%%%%%%%%%%%%%%%%
%A \rep\ $\Pi$ of $\C$ by (bounded) operators in a Hilbert space $\H$
%is called cyclic if there a -- cyclic -- vector $\Omega\in\H$ 
%such that the linear span of the set $\{\Pi(f)\Omega, \ f\in\C\}$
%is dense in $\H$. General \reps\ of $\C$ are direct sums of cyclic \reps.
%%%%%%%%%%%%%%%%%%%%%%%%%%

A positive linear functional $\omega:C(X)\to \CC$ is called a state (of the 
algebra $C(X)$) if it is normalized, i.e.~if $\|\omega\|=\omega(1)=1$.
It follows that the set of all states, or state space $E$, 
is a convex subset of the topological dual of $C(X)$. 

Cyclic representations of 
$C(X)$ are in 1-1 correspondence
with states. By the Riesz-Markov theorem, states 
%of a commutative $C^*$-algebra $\C$ 
are in
turn in 1-1 correspondence with regular normalized Borel measures on $X$.
Given such a  measure $\m$, one  has a \rep\  of
$C(X)$ by multiplication operators in the Hilbert space $L^2\bigl(X,\m\bigr)$: 
%%%%%%%%%%%%%%
\be
\label{g1}
\bigl(f\psi\bigr)(x)=  f(x)\psi(x),\ \ \psi\in 
L^2\bigl(X,\m\bigr), \ f\in C(X),
\ee
%%%%%%%%%%%%%%
%where the constant function $1\in L^2(X,\m)$ is cyclic by construction.
with corresponding state $\omega$ defined by:
%%%%%%%%%%%%%%%%%%%%%%%%%%%%%%%%%%%
\be
\label{g2}
\omega(f)=\langle 1,f1\rangle=\int_{X}f\, d\m\,.
\ee
Conversely, every cyclic \rep\ of $C(X)$ is (unitarily equivalent to) a
representation of the above  type (\ref{g1}).
Given a \rep\ Hilbert space $\H$, we will say that  a state of the form 
$f\mapsto\langle\psi,f\psi\rangle$, with a normalized $\psi\in\H$,  is a $\H$ vector state.

A \rep\ of a $C^*$-algebra is said to be faithful if it is injective, and a (regular Borel) 
measure $\m$ on a compact space  $X$ is called faithful 
if $\mu(B) \not = 0$ for all non-empty open sets $B\subset X$. It follows that a
(cyclic) \rep\ of $C(X)$ is faithful if and only if the corresponding measure $\m$
is faithful.
%i.e.~if $\int_X f d\mu > 0$ for all positive (non-zero) $f\in C(X)$. 

Let us consider the state space $E$ equipped with the weak-$\star$ topology, 
i.e.~the weakest topology such that  all complex maps 
$E\ni\omega\mapsto \omega(f)$ 
are continuous, $f\in C(X)$. The state space $E$ then becomes a compact space, 
with convergence given by:
a  net $\{\omega_{\lambda}\}_{\lambda\in I}$ converges to $\omega$ if and only if
every complex net $\{\omega_{\lambda}(f)\}_{\lambda\in I}$ converges to $\omega(f)$, 
$\forall f$.
(Nets are a generalization of sequences: a net is a family 
indexed by a directed set $I$.\footnote{A directed
set $I$ is a set equipped with a partial order relation "$\geq$" with
the property that for any $\l,\l'\in I$ there exists $\l''\in I$ such that
$\l''\geq\l,\l'$.} Only in spaces that are first countable can one 
construct  closures using sequences; 
in a general topological space a point $x$ is in the closure of a set $S$ if and only
if there is a net in $S$ converging to $x$).

The next result provides a dense set in $E$.
Let us consider the subset of Dirac states, i.e.~states
$\o_x$, $x\in X$, defined by $\o_x(f)=f(x)$. It follows that
the convex hull of Dirac states is (weak-$\star$) dense in $E$. 
%%%%%%%%%%%
%%(see e.g.~\cite[2.3.2]{BR}).
%%%%%%%%%%%%%%
Thus, given any state $\o\in E$ there is a net of 
states of the type
$$ f \mapsto \sum_{i=1}^n t_i f(x_i),$$
with $n\in\N$, $x_i\in X$, $t_i>0$, $\sum^n t_i=1$, that converges to $\o$.

%%%%%%%%%%%%%%%%%%%%%%%%%%%%%%%%%%%%%%%%%%%%%%%%
%%%%%%%%%%%%%%%%%%%%%%%%%%%%%  Section  %%%%%%%%%%%%%%%%%%%%%%%%%%%%%%%%%
\subsection{LQG basics: holonomy algebra, projective structure and uniform measure}
\label{lp}
%%%%%%%%%%%%%%%%%%%%%%%%%%%%%%%%%%%%%%%%%%%%%%%%%%%%%%%%%%%%%%%%%%%%%%%%%
Loop quantum gravity starts from the $SU(2)$ version of Ashtekar's canonical formulation of
general relativity as a special kind of gauge theory~\cite{A,B}. 
For generality and to simplify notation we consider in what follows a general connected
and compact gauge Lie group $G$.  
The classical configuration space of the theory 
is thus the space $\A$ of  $G$-connections $A$ on a principle bundle $P(\Sigma,G)$ over a 
spatial manifold $\Sigma$. For general base manifold $\Sigma$ and gauge group $G$, the bundle
$P(\Sigma,G)$ may be non-trivial (such a situation is considered in~\cite{AL3,B1} and more
recently in~\cite{LOST,OL2}). However, for the purposes of the present work it is irrelevant
whether the bundle is trivializable or not, as our results do not depend on this.
Furthermore, in the case of greatest interest, namely gravity in $3+1$ dimensions formulated
as a $G=SU(2)$ gauge theory, the bundle is actually trivial~\cite{LOST,T}.
Thus, for simplicity, we assume in what follows that the bundle $P(\Sigma,G)$ is trivial.
This allows us to identify connections $A$ with globally defined one-forms on the base manifold
$\Sigma$, with values on the Lie algebra of $G$. Moreover, one can then look at holonomies of
connections as taking values on the group $G$. 

A key 
ingredient in LQG is the particular choice of configuration 
functions, collected in a commutative $C^*$-algebra with identity called the 
ho\-lo\-no\-my algebra. The basic ($G$-valued) variables are the holonomies or
parallel transports 
%%%%%%%%%%%%%%%%%%%%%%%%%%%%
$$A\mapsto A(e):=\P \exp\bigl(-\int_e A\bigr)$$
%%%%%%%%%%%%%%%%%%%%%
along (analytically embedded oriented compact) curves $e$ in $\Sigma$, 
called edges. Basic complex functions -- called cylindrical functions --
are of the form
%%%%%%%%%%%%%%%
\be
\label{cil}
f(A)=F\left(A(e_1),\ldots,A(e_N)\right),
\ee
%%%%%%%%%%%%%%%%%%%
for arbitrary integer $N$ and  $F\in C(G^N)$.
The set Cyl of all cylindrical
functions is a commutative $*$-algebra with identity. The $C^*$-completion 
of Cyl in the
supremum norm is then the holonomy $C^*$-algebra~\cite{AI}. The spectrum of 
the algebra
is the so-called space of generalized connections $\Ab$, a compact 
extension of $\A$ that plays the role of quantum configuration space.
Following section \ref{gen}, we identify the holonomy algebra 
with 
$C(\Ab)$. It turns out that $\Ab$ coincides with a subclass of maps from 
the set
of edges of $\Sigma$ to the  group $G$. More precisely, 
each element $\bar A$ of $\Ab$
is a map $e\mapsto {\bar A}(e)\in G$, preserving the natural composition of edges~\cite{AL1,AL3,B1}. 
Conversely,
every edge $e$ defines a $G$-valued function on $\Ab$, $\bar A\mapsto  
{\bar A}(e)$. 

A quantization of configuration variables is, by construction, a 
representation of \ca. Given a measure $\m$ on $\Ab$ one thus has a 
quantization of configuration variables by multiplication operators in 
$L^2(\Ab,\m)$. In particular, cylindrical functions $f(A)$  are quantized
by the functions $f(\bar A)$.

Measure theory in $\Ab$ is well understood, due to the projective
nature of $\Ab$~\cite{AL1,AL2,AL3,B1,MM}.
One may start with the set $\Gamma$ of (appropriate) finite collections of edges, 
called graphs in $\Sigma$.
The set of graphs $\Gamma$ is naturally directed, a graph $\g'$ being 
"greater than" $\g$ ($\g'\geq\g$) if $\g$ is a subgraph of $\g'$. To each
graph $\g$ corresponds a finite dimensional configuration space $\A_\g$,
which captures the (finite number of) degrees of freedom associated to
parallel transports along the edges of $\g$. Each  $\A_\g$ 
is a compact space diffeomorphic to $G^{N_\g}$, where
the integer $N_\g$ is the number of (independent) edges in $\g$. 
There is thus an identification between $C(\A_\g)$ and $C(G^{N_\g})$,
essentially given by (\ref{cil}) above. The family 
of spaces $\{\A_\g\}_{\g\in\Gamma}$ forms a projective family, i.e.~for every pair  
$\g, \g'\in\Gamma$ such that $\g'\geq\g$,
there is a continuous surjective 
projection $p_{\g'\g}:\A_{\g'}\to \A_\g$
satisfying the consistency conditions
%%%%%%%%%%%%%
\be
\label{pr1}
p_{\g''\g}=p_{\g'\g}\, p_{\g''\g'}\, , \ \forall \g''\geq\g'\geq\g.
\ee
%%%%%%%%%%%%%%%%%%%%%%%
The  space \Ab\ is a limit -- the so-called  projective 
limit -- of the family $\{\A_{\g}\}_{\g\in\Gamma}$. In particular, this means that
there are  continuous surjective projections $p_{\g}:\Ab\to \A_\g$
satisfying the conditions
%%%%%%%%%%%%%
\be
\label{pr2}
p_{\g}=p_{\g'\g}\, p_{\g'}\, , \ \forall \g'\geq\g.
\ee
%%%%%%%%%%%%%%%%%%%%%%%
%%%%%%%%%%%%%%%%%%%%%%%
Notice that the $*$-algebra of functions on \Ab\ of the form $p^*_{\g}f$, $f\in C(\A_\g)$, $\g\in\Gamma$,
where $p^*_{\g}$ denotes pull-back, is naturally isomorphic to the algebra Cyl of cylindrical functions.
As usual, we will not distinguish between the two algebras.

It follows from the above structure~\cite{AL2} that 
(normalized regular Borel) measures on \Ab\  are in 1-1
correspondence with families $\{\m_\g\}_{\g\in\Gamma}$ of 
measures
$\m_\g$ on the spaces $\A_\g$, satisfying the consistency conditions:
%%%%%%%%%%%%%%%%%%%
\be 
\label{pr5}
\int_{\A_{\g'}} p^*_{\g'\g}fd\m_{\g'}=\int_{\A_{\g}} fd\m_{\g},\ 
\forall \g'\geq\g.
\ee
%%%%%%%%%%%%%%%%%%%%%%%%%
The correspondence between a measure $\m$ on $\Ab$ and the associated family 
$\{\m_\g\}_{\g\in\Gamma}$ is given by
%%%%%%%%%%%%%%%%%%%
\be 
\label{pr7}
\int_{\Ab} p^*_{\g}fd\m =\int_{\A_{\g}} fd\m_{\g}, \ \forall \g,\ 
\forall f \in C(\A_\g).
\ee
%%%%%%%%%%%%%%%%%%%%%%%%%%
As a counterpart of the projective structure of \Ab, every  Hilbert
space $\H=L^2(\Ab,\m)$ acquires a so-called inductive structure, as follows (see~\cite{T} for further details).
Let $\H_\g$  denote the Hilbert space $L^2(\A_\g,\m_\g)$. Due to (\ref{pr5}), the pull-back's
$p^*_{\g'\g}$, $\g'\geq\g$, define injective isometries 
$p^*_{\g'\g}:\H_\g \to \H_{\g'}$ satisfying
consistency conditions following from (\ref{pr1}). Likewise, it follows from 
(\ref{pr7}) that the pull-back's 
$p^*_{\g}$ define transformations
$p^*_{\g}:\H_\g\to \H$, 
which are unitary when considered as maps 
%%%%%%%%%%%%%%%%%
\be
\label{z3}
p^*_{\g}:\H_\g\to p^*_{\g}\H_{\g}
\ee
%%%%%%%%%%%%
onto their images, the closed subspaces $p^*_{\g}\H_{\g}$. The linear maps
$p^*_{\g}$ and $p^*_{\g'\g}$ are related by consistency conditions corresponding to (\ref{pr2}), namely:
%%%%%%%%%%%%%%%%%%%
\be
\label{z2}
p^*_{\g}=p^*_{\g'}p^*_{\g'\g},\ \forall \g'\geq \g\, .
\ee
%%%%%%%%%%%%%%%%%%%%%%%%%
Notice that the subspace ${\rm Cyl}=\bigcup_{\g\in\Gamma}p^*_{\g}C(\A_\g)$, and therefore the reunion
$\bigcup_{\g\in\Gamma}p^*_{\g}\H_\g$, is dense in $\H$. This fact will be used repeatedly 
in section \ref{pro}.

The \rep\ of \ca\ upon which LQG is actually developed is based on the  
Ashtekar-Lewandowski, or uniform measure $\m_0$~\cite{AL1}. 
The measure $\m_0$ is defined by a family
$\{{\m_0}_\g\}_{\g\in\Gamma}$ such that, for each $\g$, ${\m_0}_\g$
is the
image of the Haar measure on $G^{N_\g}$, under the natural 
identification $\A_\g\equiv G^{N_\g}$. The corresponding Hilbert space is the so-called
kinematical Hilbert space $\H_0:=L^2(\Ab,\m_0)$. The measure $\m_0$ has
remarkable and unique properties, allowing a
quantization of the full LQG
kinematical algebra together with implementations of the constraints of general relativity
(see~\cite{F,LOST,OL,OL2,ST1} for uniqueness results concerning the quantization of flux variables together 
with the implementation of the diffeomorphism 
constraint,~\cite{T,Tpho} for a discussion on the implementation of the hamiltonian constraint,
and~\cite{AL5,R,T} for general treatments of LQG).

In what concerns us here, only one property of $\m_0$ will be used, namely its faithfulness~\cite{AL1,AL2},
which, as seen before, is the same as faithfulness of the corresponding \rep\ of \ca.

%%%%%%%%%%%%%%%%%%
%%%%%%%%%%%%%%%%%%%%%%%%%%%%%  Section  %%%%%%%%%%%%%%%%%%%%%%%%%%%%%%%%%
\section{Denseness of $\H_0$ vector states}
\label{newgen}
%%%%%%%%%%%%%%%
It is argued in~\cite{T2,T1} that the $\m_0$ \rep\ is in some
sense a "fundamental \rep" of the algebra \ca, meaning that new \reps\ can be obtained from
the $\H_0$ inner product. 
Using results from section \ref{gen} and well known separation 
properties of 
compact
Hausdorff spaces, we will now show that every \ca\ state can indeed be obtained,
as a weak-$\star$ limit, from $\H_0$ vector states.
This is in fact a very general result, relying on faithfulness only.

Let us start by constructing a directed set needed in what follows.
%%%%%%%%%%%%%%%%
Let $\{\bar A_1,\ldots,\bar A_n\}$ be a finite set of distinct points of \Ab\ and
consider a set of disjoint  open sets $\{U_1,\ldots,U_n\}$,
with $\bar A_i\in U_i$.
% (such a set exists due to the separation properties of $X$.)
Then the set of ordered n-tuples of the form $(B_1,\ldots,B_n)$
with $B_i\subset U_i,\ \bar A_i\in B_i$, $B_i$ open, is clearly a directed
set with respect to the following partial order relation: 
$$ (B_1,\ldots,B_n) \geq (D_1,\ldots,D_n)\ \ {\rm whenever} 
\ \ B_i\subset D_i\ \forall i.$$
%%%%%%%%%%%%%%%%
%{\bf proof:} The set of such n-tuples is obviously non empty.
%It is trivial to see that the order relation is well defined. 
%It is also trivial
%to show that the set is directed, since 
%the intersection of two 
%open neighborhoods of a given point is again a open neighborhood  $\Box$.
%%%%%%%%%%%%%%% 
%Let $\m_0$ be a faithful measure on $X$,
%$\langle\, ,\, \rangle_0$ be the associated $L^2$ inner product
%and $\H_0:=L^2(X,\m_0)$ the corresponding Hilbert space. 
Let then $E(\H_0)$ be the set of $\H_0$ vector states, i.e.~the set of states 
of the algebra \ca\ 
defined by
normalized vectors $\psi\in \H_0$ as follows:
$$\ca\ni f \mapsto \langle\psi,f\psi\rangle_0,$$
where $\langle\ ,\rangle_0$ denotes the $\H_0$ inner product.
We will now show that
the convex hull of Dirac states lies in the weak-$\star$ closure of $E(\H_0)$.
%%%%%%%%%%%%%%%%%%%
To prove it, let us fix a state in the convex hull of Dirac states: 
$$f\mapsto \sum^n t_i f(\bar A_i),\  \
\ n\in\N,\ \bar A_i\in\Ab,\ t_i>0, \  \sum^n t_i=1.$$
Consider then disjoint open sets $\{U_1,\ldots,U_n\}$ in $\Ab$,
with $\bar A_i\in U_i$, and 
let $I$ be the directed set $\{(B_1,\ldots,B_n)\}$
defined as above, with $\bar A_i\in B_i\subset U_i$.
Let $\chi_{B_i}$ be the characteristic function of the open set $B_i$.
Clearly $\|\chi_{B_i}\|_0^2=\mu_0(B_i)\not =0$, since $\m_0$ is faithful. 
Let 
$$\psi_{B_i}:={\chi_{B_i} \over \|\chi_{B_i}\|_0} \in \H_0$$
and consider still
$$\psi_{B_1\ldots B_n}:=\sum^n t_i^{1/2}\psi_{B_i}.$$
We also have 
$$\|\psi_{B_1\ldots B_n}\|_0^2=1,$$
since the sets $B_i$ are disjoint and $\sum t_i=1$.
Consider finally the net of states indexed by $I$:
$$f\mapsto \langle \psi_{B_1\ldots B_n}, f\, \psi_{B_1\ldots B_n}\rangle_0=
\sum^n {t_i\over \mu_0(B_i)}\int_{B_i}fd\mu_0\ , \ (B_1,\ldots, B_n)\in I.$$
One can now show that $\langle \psi_{B_1\ldots B_n}, f\, 
\psi_{B_1\ldots B_n}\rangle_0$ converges, $\forall f$, to the given value
$\sum t_i f(\bar A_i)$. In order to do this,
let us  fix $f$ and consider any $\epsilon >0$. Let $\epsilon' >0$
be such that $n\epsilon'<\epsilon$. Since $f$ is continuous, for each $\bar A_i$
there is an open set $V_i\ni \bar A_i$ such that
$$|f(\bar A)-f(\bar A_i)|<\epsilon'\ \ \forall \bar A\in V_i.$$
In particular, there are open sets $B^0_i,$ 
$$B^0_i:=V_i\cap U_i\subset U_i\ ,\ \bar A_i\in B^0_i,$$
such that
$$|f(\bar A)-f(\bar A_i)|<\epsilon'\ \ \forall \bar A\in B^0_i.$$
Let us take 
$$I\ni (B_1,\ldots,B_n)\geq (B^0_1,\ldots,B^0_n).$$
Then
$$|\langle \psi_{B_1\ldots B_n}, f\, \psi_{B_1\ldots B_n}\rangle_0 -
\sum^n t_i f(\bar A_i) |= |\sum^n {t_i\over \mu_0(B_i)}\int_{B_i}(f-f(\bar A_i))
d\mu_0|\leq$$
$$\leq\sum^n sup_{B_i} |f(\bar A)-f(\bar A_i)|\leq n\epsilon'<\epsilon\ ,
\forall (B_1\ldots B_n)\geq(B^0_1\ldots B^0_n),$$
%%%%%%%%%%%%%%%
showing that the convex hull of Dirac states is in the  closure of $E(\H_0)$.

Combining the above result  with the denseness of the convex hull of Dirac
states (section \ref{gen}) one then concludes that the set
%%%%%%%%%%%%%%%%%%%%
%\begin{prop}
%\label{c1}
%Let $X$ be a compact Hausdorff space and 
%$\mu_0$ a regular normalized faithful Borel measure on $X$.
$E(\H_0)$  is weak-$\star$ dense in the space $E$ of all states of \ca.
%\end{prop}
%%%%%%%%%%%%%%
Thus, given any state $\omega$ of  $C(\Ab)$, there is a 
net of vectors $\psi_{\l}\in\H_0$,  where $\lambda$ belongs to some 
directed set $I$, such that the net
$\langle \psi_{\l},f\psi_{\l}\rangle_0$ converges to 
$\omega(f)$, $\forall f\in C(\Ab)$.
In particular, given
any (regular  
Borel) measure $\m$ and $\psi\in L^2(\Ab,\m)$,  
there  is a \h0\ net $\{\psi_{\lambda}\}$
such that
$\langle\psi_{\lambda},f\psi_{\lambda}\rangle_0$ converges to 
$\langle\psi,f\psi\rangle:=\int f |\psi|^2d\mu$, $\forall f$.

%%%%%%%%%%%%%%%%%%%%%%%
%%%%%%%%%%%%%%%%%%%%%%%%%%%%%  Section  %%%%%%%%%%%%%%%%%%%%%%%%%%%%%%%%%
\section{Mapping  vector states to \h0\ cylindrical nets}
\label{pro}
%%%%%%%%%%%%%%%%%%%%%%%%%%%%%%%%%%%%%%%%%%%%%%%%%%%%%%%%%%%%%%%%%%%%%%%%%
Despite major progresses in the LQG programme, the identification 
of  semiclassical or low energy states has proved very hard,
and this task is not yet completed. Moreover, naturally constructed candidate semiclassical
states are typically not elements of \h0, but of some extension thereof.
As extensions of \h0\ are already required for different reasons, e.g.~in order 
to solve the diffeomorphism constraint~\cite{ALMMT,MTV}, it is perhaps not
surprising that semiclassical analysis leads us to consider
extensions of \h0\ as well. It does, however, create difficulties concerning the
interpretation of candidate states, since well defined quantum operators that
ultimately can give meaning to those states are defined in \h0, and not,
{\it a priori}, in the required extensions.

Two types of "generalized states" occur naturally in relation to LQG semiclassical analysis,
namely  complex (not necessarily continuous) linear functionals over the space Cyl of cylindrical functions,
and states of \ca\ that are not realizable as \h0\ vector 
states~\cite{AG,AL4,Bo,T,T1,Va1,Va2,Va4,Ve}. 
Although there is a very interesting interplay between these two types of 
objects~\cite{AL4,T1,Va2}, the relation between them  is not fully
clear in general.

The way in which  linear functionals $\Psi$ -- also called distributions in this context --  
are dealt with in LQG is to trade 
them for corresponding graph-labelled nets $\{p^*_\g\Psi_\g\}_{\g\in\Gamma}$ of \h0\ 
vectors~\cite{AL4,T1}. The mapping from distributions to \h0\ nets is achieved
through the so-called cut-off procedure, as follows. 
The cut-off, 
up to a graph $\g$, of a linear functional $\Psi$ over Cyl is 
defined (when it exists) as the unique $\Psi_\g \in {\H_0}_\g$ such that 
${\langle{ \bar  \Psi}_\g,f\rangle_0}_\g=\Psi(p^*_\g f)$ is satisfied $\forall
f\in C(\A_\g)$, where ${\H_0}_\g:=L^2(\A_\g,{\m_0}_\g)$ and ${\langle\, ,\rangle_0}_\g$ denotes the
corresponding inner product.
The nets $\{p^*_\g\Psi_\g\}_{\g\in\Gamma}$ typically do not converge in the \h0\ norm.
Nevertheless, important 
properties of the original distribution $\Psi$ are  captured  in a weaker
sense~\cite{AL4,T1}.

The purpose of the present section is to point out that methods similar to 
the cut-off procedure  can also be applied to an important class of states of \ca,
and to study convergence properties of such a generalized cut-off construction.

In particular, we consider
vector states corresponding to some Hilbert space 
$\H=L^2(\Ab,\m)$, with the requirement that the measure $\m$ is such
that each measure $\m_\g$ of the associated family is absolutely continuous
with respect to ${\m_0}_\g$. While this is not the
general case, it covers the most potentially interesting situations.
The point is simply that ${\m_0}_\g$ is (essentially) 
the uniform Haar measure on $G^{N_\g}$, and therefore  more general 
measures would give rise to representations that look pathological already at 
the finite dimensional level, since they assign non-zero measure values to
subsets of $G^{N_\g}$ of zero Haar measure.\footnote{This is reminiscent of the generic situation 
in quantum field theory, where interaction measures are typically 
equivalent  to the Lebesgue measure
when restricted to finite dimensions. Moreover, one could probably safely assume that $\m_\g$
is in fact equivalent to ${\m_0}_\g$, $\forall \g$, i.e.~that ${\m_0}_\g$ is in turn 
continuous with respect to $\m_\g$. We work with the weaker condition
for generality, as it poses no further difficulties.}

Let us then consider a measure
$\m$ on $\Ab$,  defined by a family
$\{\m_\g\}_{\g\in\Gamma}$ such that there is a family of
positive functions
$R_\g\in L^1(\A_\g,{\m_0}_\g)$ satisfying $d\m_\g=R_\g d{\m_0}_\g$, 
$\forall \g$. Let $\H$ be the corresponding Hilbert space $L^2(\Ab,\m)$.
The non-trivial situation occurs when the Radon-Nikodym 
derivative $d\m/d\m_0$ fails to exist, so that the natural inner product 
preserving transformation 
%%%%%%%%%%%%%%%%%
\be
\label{map}
\H\ni\psi\mapsto (d\m/d\m_0)^{1/2}
\psi\in\H_0
\ee
%%%%%%%%%
is not available. In fact, when $d\m/d\m_0$ exists the transformation (\ref{map}) maps $\H$
to a subspace $\H^{\m}_0\subset\H_0$ of functions supported on the support of $\m$, and 
transforms the $\m$ \rep\ of \ca\ into the restriction to $\H^{\m}_0$ of the $\m_0$ \rep.
If, moreover, $d\m_0/d\m$ is also defined, then (\ref{map}) establishes a unitary equivalence
between \reps.

In our case, due to the existence of $d\m_\g/d{\m_0}_\g$, $\forall \g$, 
one has a family of inner product preserving maps from 
$\H_\g=L^2(\A_\g,\m_\g)$ to ${\H_0}_\g$. Combining these maps with orthogonal projections
on the $\H$ side and pull-back's on the \h0\ side, one can then define a weak 
version of the transformation (\ref{map}),
consisting of a map from $\H$ to graph-labelled nets in $\H_0$.

Let then  
%%%%%%%%%%%%%%%%
\be
\label{01}
P_\g:\H\to p^*_{\g}\H_{\g}
\ee
%%%%%%%%%%%%%%
denote the orthogonal projection onto 
the closed subspace $p^*_{\g}\H_{\g}$.
Notice that  consistency condition (\ref{z2}) leads to
$p^*_\g\H_\g\subset p^*_{\g'}\H_{\g'}$, $\forall \g'\geq\g$, and therefore $P_{\g'}$
is the identity operator on every subspace $p^*_\g\H_\g$ such that $\g'\geq\g$.
Let us also introduce the composition $\pi_\g:=(p^*_{\g})^{-1}P_\g$, 
$\pi_\g:\H\to \H_\g$. It is easily seen that $\pi_\g$ satisfies
%%%%%%%%%%
\be
\label{02}
\langle \pi_\g\psi,f\rangle_\g=\langle \psi,p^*_\g f\rangle,\ \forall
f\in\H_\g,
\ee
%%%%%%%%%%%%%%%%%%
where $\langle\, ,\rangle_\g$ and  $\langle\, ,\rangle$ denote inner
products in $\H_\g$ and $\H$, respectively. Notice also that nothing is lost 
at this stage when trading a given $\psi\in\H$ by the family $\{\pi_\g\psi\}_{\g\in\Gamma}$, 
since the net of $\H$ vectors 
$\{p^*_{\g}\pi_\g\psi\}_{\g\in\Gamma}=\{P_\g\psi\}_{\g\in\Gamma}$
clearly converges to $\psi$.

One can now bring in the multiplication operators $R_\g^{1/2}$, obtaining a
map  from $\H$ to \h0\ nets as follows.
Each $\psi\in\H$ is mapped to the net $\{\psi_\g\}_{\g\in\Gamma}$, where
$\psi_\g\in\H_0$ is defined by
%%%%%%%%%%%%%%%%%
\be
\label{03} 
\psi_\g=p_\g^*(R_\g^{1/2}\pi_\g\psi).
\ee
%%%%%%%%%%%%%%%%%%
To see in what sense the map (\ref{03}) generalizes (\ref{map}), 
%%%
%$\psi\mapsto (d\m/d\m_0)^{1/2}\psi$, 
let us first show that the image 
$(d\m/d\m_0)^{1/2}\psi$ 
is recovered from the net $\{\psi_\g\}_{\g\in\Gamma}$ in the trivial
case in which  $d\m/d\m_0$ actually exists. More precisely, we will show that if there exists
$R\in L^1(\Ab,\m_0)$ such that
%%%%%%%%%%%%%%%%
$d\m=R d\m_0$,
then, $\forall\psi\in\H$, the net $\{\psi_\g\}_{\g\in\Gamma}$
 converges in the $\H_0$ norm to $R^{1/2}\psi$.
%%%%%%%%%%%%%%%%%%%%%%%%%%

To prove the above we  use the following  facts:
{\it i\/}) the set Cyl of cylindrical functions is dense in $\H$, and
{\it ii\/}) the net $\{p^*_\g R^{1/2}_\g\}_{\g\in\Gamma}$ converges, in
the $\H_0$ norm, to $R^{1/2}$~\cite[B.2.7]{Ya}.
Let us then fix arbitrary $\psi\in\H$ and $\epsilon > 0$. Let us choose
$\g_0$ and $f\in C(\A_{\g_0})$ such that $\|p^*_{\g_0}f-\psi\|<\epsilon/6$,
where $\|\, \|$ denotes the $\H$ norm.
Also, let us choose $\g_1$ and $\g_2$ such that $\|P_\g\psi-\psi\|<\epsilon/9$, 
$\forall \g\geq\g_1$, and 
$\|p^*_\g R^{1/2}_\g-R^{1/2}\|_0<\epsilon/(3\|p^*_{\g_0}f\|_{C^*})$,
$\forall \g\geq\g_2$, where $\|\, \|_{C^*}$ denotes the \ca\ $C^*$-algebra norm. 
Moreover, let $\g_3$ be such that $\g_3\geq \g_0,\g_1,\g_2$. We now have
%%%%%%%%%%%%%%%
\ba
\label{05}
\|\psi_\g-R^{1/2}\psi\|_0&=& \|(p^*_\g R^{1/2}_\g-R^{1/2})P_\g\psi+
R^{1/2}(P_\g\psi-\psi)\|_0\leq \nonumber \\
&\leq& \|(p^*_\g R^{1/2}_\g-R^{1/2})P_\g\psi\|_0+\|P_\g\psi-\psi\|,
\ea
%%%%%%%%%%%%%%%%%%
and
%%%%%%%%%%%%%%%%%%
\ba
\label{06}
&{}&\|(p^*_\g R^{1/2}_\g-R^{1/2})P_\g\psi\|_0= \nonumber \\
&{}&\|(p^*_\g R^{1/2}_\g-R^{1/2})(P_\g\psi-p^*_{\g_0}f) +
(p^*_\g R^{1/2}_\g-R^{1/2})p^*_{\g_0}f\|_0\leq \nonumber \\ 
&{}&\|(p^*_\g R^{1/2}_\g-R^{1/2})(P_\g\psi-p^*_{\g_0}f)\|_0 +
\| p^*_{\g_0}f\|_{C^*}\|p^*_\g R^{1/2}_\g-R^{1/2}\|_0.
\ea
%%%%%%%%%%%%%%%%%%%%%%%%%%%%%
Regarding the first term in the last line of (\ref{06}), notice that
$P_\g\psi-p^*_{\g_0}f$
is an element of
$p^*_\g\H_\g$, $\forall \g\geq\g_0$, and that both $p^*_\g R_\g^{1/2}$ and $R^{1/2}$ are bounded
operators of unit norm, from $p^*_\g\H_\g$ to $\H_0$.
Thus
%%%%%%%%%%%%%%%%%%
\ba
\label{07}
\|(p^*_\g R^{1/2}_\g-R^{1/2})(P_\g\psi-p^*_{\g_0}f)\|_0&\leq&
2\|P_\g\psi-p^*_{\g_0}f\|\leq \nonumber \\
&\leq& 2\|P_\g\psi-\psi\|+2\|\psi-p^*_{\g_0}f\|.
\ea
%%%%%%%%%%%%%%%%%%
Putting (\ref{05}), (\ref{06}) and (\ref{07}) together we get, $\forall \g\geq\g_3$:
$$\|\psi_\g-R^{1/2}\psi\|_0\leq 3\|P_\g\psi-\psi\|+2 \|\psi-p^*_{\g_0}f\|+
\| p^*_{\g_0}f\|_{C^*}\|p^*_\g R^{1/2}_\g-R^{1/2}\|_0 < \epsilon\, ,
$$
thus concluding the proof.

Returning to the general case, we 
%will 
now establish the exact sense
in which  (\ref{03}) generalizes the transformation (\ref{map}). Specifically we  show
that, $\forall \psi',\psi\in\H$ and $\forall F\in \ca$, 
%%%%%%%%%%%%%%%%%%%%%%%%%
the complex net $\{\langle \psi'_\g,F\psi_\g\rangle_0\}_{\g\in\Gamma}$
converges to $\langle \psi', F\psi\rangle$.
%%%%%%%%%%%%%%%%%%%%%%%%

We will use the fact that the set of 
cylindrical functions
%%%%%%%%
%%, i.e.~functions of the type $p_\g^*f$, for some
%%$\g\in\Gamma$ and $f\in C(\A_\g)$, 
%%%%%%%%%%%
is dense in \ca.  
We  start precisely by proving the result for cylindrical $F$,
using (\ref{z2}) and unitarity of $p^*_\g$ (\ref{z3}).
%
%, the definition of $\pi_\g$ and the condition $d\m_\g=R_\g d{\m_0}_\g$.
%
Let us then fix $\psi',\psi\in\H$, $\g_0\in\Gamma$, $f\in C(\A_{\g_0})$ and 
take $F=p_{\g_0}^*f$. For every
$\g\geq\g_0$ we have 
%%%%%%%%%%%%%%%%%%%%%%%%%
\ba
\label{pr9}
\langle \psi'_\g,(p_{\g_0}^*f) \psi_\g\rangle_0 & = & 
{\langle R_\g^{1/2}\pi_{\g}\psi',(p_{\g\g_0}^*f) 
R_\g^{1/2}\pi_{\g}\psi\rangle_0}_\g= \nonumber \\
\langle \pi_{\g}\psi',(p_{\g\g_0}^*f) \pi_\g \psi\rangle_\g & = &
\langle P_{\g}\psi',(p_{\g_0}^*f) P_{\g}\psi\rangle = \nonumber \\
\label{pr10}
& = & \langle\psi',(p_{\g_0}^*f) P_{\g}\psi\rangle,
\ea
%%%%%%%%%%%%%%%%%%%%%%
where in the last line we have used the fact that
$(p_{\g_0}^*f) P_{\g}\psi$ is actually 
an element of
$p_{\g}^*\H_{\g}$. 
Thus
%%%%%%%%%%%%%%%%%%%
%\be
%\label{pr13}
$$|\langle \psi'_\g,(p_{\g_0}^*f)\psi_\g\rangle_0
-\langle\psi',(p_{\g_0}^*f)\psi\rangle|=
|\langle\psi',(p_{\g_0}^* f)(P_{\g}\psi-\psi)\rangle|\leq
\|\psi'p_{\g_0}^*\bar f\| \|P_{\g}\psi-\psi\|.
$$
%\ee
%%%%%%%%%%%%%%%%
Convergence now follows, since $\psi'p_{\g_0}^*\bar f$ is fixed.
We turn next to general $F$, using the denseness of cylindrical functions
and the continuity of the \reps\ of \ca. 
Let us then fix $\psi',\psi\in\H$,  $F\in \ca$, $\epsilon>0$ and 
choose $\g_0$ and $f\in C(\A_{\g_0})$ such that
%%%%%%%%%%%%
$$\|p_{\g_0}^*f-F\|_{C^*}<\epsilon/(3\|\psi'\|\|\psi\|)\, .$$
%%%%%%%%%%%%
Let $\g_1\geq\g_0$ be such that
%%%%%%%%%%%%%%%%%%%%
$$|\langle \psi'_\g,(p_{\g_0}^*f)\psi_\g\rangle_0
-\langle\psi',(p_{\g_0}^*f)\psi\rangle|< \epsilon/3,\ \forall \g\geq\g_1.$$
%%%%%%%%%%%%%%%%%%%%
The proof can now be completed, since $\forall\g\geq\g_1$ we find
%%%%%%%%%%%%%%%%%%%%%%
\begin{eqnarray*}
&\!&|\langle \psi'_\g,F\psi_\g\rangle_0-\langle\psi',F\psi\rangle|= \nonumber \\
&\!&|\langle \psi'_\g,(F-p_{\g_0}^*f)\psi_\g\rangle_0+
\langle\psi',(p_{\g_0}^*f-F)\psi\rangle+
\langle \psi'_\g,(p_{\g_0}^*f)\psi_\g\rangle_0
-\langle\psi',(p_{\g_0}^*f)\psi\rangle| \nonumber \\
&\!&<\|p_{\g_0}^*f-F\|_{C^*}
\bigl(\|\psi'_\g\|_0\|\psi_\g\|_0+\|\psi'\|\|\psi\|\bigr)
+\epsilon/3 \nonumber\\
&\!&=\|p_{\g_0}^*f-F\|_{C^*}
\bigl(\|P_\g\psi'\|\|P_\g\psi\|+\|\psi'\|\|\psi\|\bigr)
+\epsilon/3 \, <\, \epsilon\, .
\end{eqnarray*}
%%%%%%%%%%%%%%%%%%%%%%%%%
The above results  support the viewpoint of considering
the map (\ref{03})  as a natural weak 
substitute for transformation (\ref{map}) (for the measures under 
consideration), thus generalizing the cut-off procedure in what concerns 
mapping elements of $\H$ into graph-labelled nets in \h0.

\section*{Acknowledgments}
\noindent 
I am most grateful to Jos\'e Mour\~ao. I would also like to thank Thomas Thiemann. 
This work was supported in part by 
%projects 
POCTI/33943/MAT/2000,
%, CERN/P/FIS/43171/2001, 
POCTI/FNU/\-49529/2002 
and POCTI/FP/FNU/50226/2003.
%%%%%%%%%%%%%%%%%%%%%%%%%%%%%%%%
%\newpage 

%%%%%%%%%%%%%% Bibliography %%%%%%%%%%%%%%%%%%%%%%%

%\section*{References}
%\bigskip

%%%%%%%%%%%%%%%%%%%%%%%%%%%%%%%%%%

%%%%%%%%%%%%%%%%%%%%%%%%%%%%%%%%%

\end{document}